\documentclass[useAMS,usenatbib, referee]{mn2e}
\usepackage{graphicx}
\usepackage{graphicx}
\title[The  explosive end of Pop-III AGB stars]{An Explosive End to Intermediate-Mass Zero-Metallicity Stars and Early Universe Nucleosynthesis}
\author[H.B. Lau, R.J. Stancliffe \& C.A. Tout]{Herbert H.B. Lau\thanks{E-mail:
HBL21@ast.cam.ac.uk}, Richard J. Stancliffe  and Christopher A. Tout\\
Institute of Astronomy, The Observatories, Madingley Road, Cambridge CB3 0HA}
\begin{document}
\bibliographystyle{mn2e}
\date{Accepted 0000 December 00. Received 0000 December 00; in original form 0000 October 00}

\pagerange{\pageref{firstpage}--\pageref{lastpage}} \pubyear{0000}

\maketitle

\label{firstpage}

\begin{abstract}
We use the Cambridge stellar evolution code \textsc{stars} to model the evolution of $5\,M_\odot$ and $7\,M_\odot$ zero-metallicity stars. With enhanced resolution at the hydrogen and helium burning shell in the AGB phases, we are able to model the entire thermally pulsing asymptotic giant branch (TP-AGB) phase. The helium luminosities of the thermal pulses are significantly lower than in higher metallicity stars so there is no third dredge-up. The envelope is enriched in nitrogen by hot-bottom burning of carbon that was previously mixed in during second dredge-up. There is no $s$-process enrichment owing to the lack of third dredge up. The thermal pulses grow weaker as the core mass increases and they eventually cease. From then on the star enters a quiescent burning phase which lasts until carbon ignites at the centre of the star when the CO core mass is $1.36\,M_\odot$. With such a high degeneracy and a core mass so close to the Chandrasekhar mass, we expect these stars to explode as type 1.5 supernovae,  very similar to Type Ia supernovae but inside a hydrogen rich envelope. 
\end{abstract}


\section{introduction}
The primordial generation of stars, commonly referred to as Population-III stars or zero-metallicity stars, should have the composition of the interstellar (ISM) just after Big Bang nucleosynthesis and hence have a negligible abundance of metals. It has been a popular belief that, in the absence of heavy elements and dust grains, cooling mechanisms are inefficient and favour the formation of massive or very massive stars. The standard mechanism, that accretion is terminated by radiation pressure on dust grains \citep[e.g.][]{Wolfire} in metal-rich gas, is not effective for gas with a primordial composition because there is no dust. Recently, it has been speculated that accretion could instead be turned off through the formation of an H-II region \citep{Omukai2} or through the radiation pressure exerted by trapped Ly$\alpha$ photons \citep{tan}. It has also been shown \citep*{Palla,Yoshii} that even a small fraction of molecular hydrogen can provide a significant contribution to cooling via rotational and vibrational transitions. The resulting Jeans mass of a pure H and He cloud could then be relatively small and may even fall below $0.1\,M_\odot$. Owing to this complexity and our lack of understanding of star formation, the initial mass function (IMF) of zero-metallicity stars remains uncertain. Using one- and two-dimensional hydrodynamical simulations, \shortcite{Nakamura} showed that, depending on the initial density of the filamentary primordial gas cloud, there is an alternative result of the fragmentation of primordial filaments. For high density gas clouds, because the $H_{2}$ cooling is more effective owing to three-body reactions, filaments can contract and the fragmentation mass can be lowered to $1\,M_\odot$. Hence, they suggested a bimodal IMF with peaks close to $1\,M_\odot$ and $100\,M_\odot$. \shortcite{Johnson} have recently suggested that the formation of primordial low- and intermediate-mass stars is viable. It is likely that extremely low- to zero-metallicity AGB stars did form in the early Universe and so their evolution and contribution to the nucleosynthesis history should be investigated. 

In this paper, we describe the final stages of the evolution of $5\,M_\odot$ and $7\,M_\odot$ zero-metallicity stars, in particular the TP-AGB phase, and show that type 1.5 supernova are the probable fate of these stars. This shows that supernova explosions at zero-metallicity do not require initial stellar masses as high as in the case of solar metallicity and this can lead to important implications for galactic chemical evolution. 

The idea of a type 1.5 supernova or type I$\frac{1}{2}$ supernova is not new. Such supernova occur if a star's degenerate carbon/oxygen core grows up to near Chandrasekhar mass before it loses its envelope. This possibility was suggested, for instance, by \shortcite{Arnett}, \shortcite{Iben},  \shortcite{Willson} and recently by \shortcite{Zijlstra}. As the core mass approaches $1.38\,M_\odot$ carbon ignites and the thermal runway in degenerate material cannot be delayed long enough to prevent an explosion from disrupting the entire star. Because the exploding star is a red supergiant with a hydrogen-rich envelope, its spectrum and early light curve should closely resemble that of a supernova of type II. However, a substantial amount of radioactive Ni and Co is liberated by the exploding core, thus producing a late exponential luminosity decline which could look like a  type Ia supernova \citep{Iben}. 

Lower metallicity stars have weaker stellar winds and thus their degenerate cores are able to grow up to near the Chandrasekhar mass and carbon ignition can lead to thermonuclear runaway and explosion. The mass-loss rate of low-metallicity AGB objects is uncertain but it is highly probable that it is lower than at solar metallicity. A faster core growth rate also increases the possibility of supernova Type 1.5. \shortcite*{Gilpons} also reached the similar conclusion that supernovae of Type~1.5 are inevitable in the evolution of zero-metallicity stars between $5\,M_\odot$ and $7\,M_\odot$, based on estimates of the mass-loss rate and core-growth rate. However, they did not compute the full evolution and estimated the two rates from the first few pulses. In this paper, we describe the full computation of $5\,M_\odot$ and $7\,M_\odot$ models and show that carbon ignition at the degenerate core does occur. 

\section{The STARS code}
We use the Cambridge stellar evolution code \textsc{stars} to model the evolution of primordial intermediate-mass stars. It was originally written by \shortcite{Eggleton} and has been updated by many authors \citep{Pols}.  \shortcite{STARSopacity} updated the opacity tables to use the latest OPAL calculations \citep{Iglesias} and these new tables also account for changes in opacity with variations in the carbon and oxygen abundances. We use their zero-metallicity opacities in this work (Eldridge, private communication based upon calculations by \citealt{Ferguson}). Unlike most codes which treat mixing in a separate step, this code solves the equations of stellar structure, nuclear burning and mixing simultaneously. 

 Another unique feature of the \textsc{stars} code is its use of a self-adaptive non-Lagrangian, non-Eulerian mesh. The mesh adapts so that mesh points concentrate in the physically important regions, such as burning shells and ionization zones, where things are changing most rapidly. During the asymptotic giant branch (AGB) phase, more mesh points are needed in the hydrogen and helium burning regions in order to resolve interaction between the two shells. Failure to resolve these regions properly can result in non-convergence of models or erroneous results. For example, see \shortcite{Straniero95} for a description of what happens with insufficient resolution. We use the AGB mesh spacing function described by \shortcite*{Stancliffe} to resolve thermal pulses. In the models presented here 999 meshpoints were used. Such a large number of mesh points is needed to avoid resolution problems, not only during the AGB phase but also at the end of helium burning and to avoid the numerical problem for zero-metallicity stars described by \shortcite{Tartu}. We do not include convective overshooting at any stage of the evolution. The inclusion of convective overshooting could lead to a larger core mass at the end of core helium burning but is unlikely to qualitatively change the evolution, though the surface abundances of metals could increase \citep{Gilpons}. We assume there is no mass loss from the star. The mixing-length parameter \citep{Bohm} $\alpha$ is 1.925 based on calibration to a solar model. The helium mass fraction is chosen to be 0.25 to reflect the prediction of primordial helium abundance from the observed deuterium abundance, baryon density and a spectroscopic sample of extragalactic H-II regions \citep{Fukugita}.

\section{early evolutionary phases of the $7\,M_\odot$ model}
The evolution of our $7\,M_\odot$ zero-metallicity star differs significantly from higher metallicity stars because of the absence of carbon, nitrogen and oxygen. Hydrogen cannot be burned through the CNO cycle, so it is burned via the proton-proton chain only. This chain is much less temperature dependent, so zero-metallicity stars are considerably hotter than their higher metallicity counterparts and their main-sequence lifetimes are much shorter. At the start of the main sequence, a convective core is driven by the pp-chain. The core ceases to be convective while hydrogen is still abundant. The temperature rises and becomes hot enough that carbon is produced by the triple-$\alpha$ reaction before hydrogen is exhausted. Because carbon is present in a hydrogen-rich region, the CNO cycle can now take place and drive a convective region in the core again.

\begin{figure} 
\includegraphics [angle=0, width =\columnwidth] {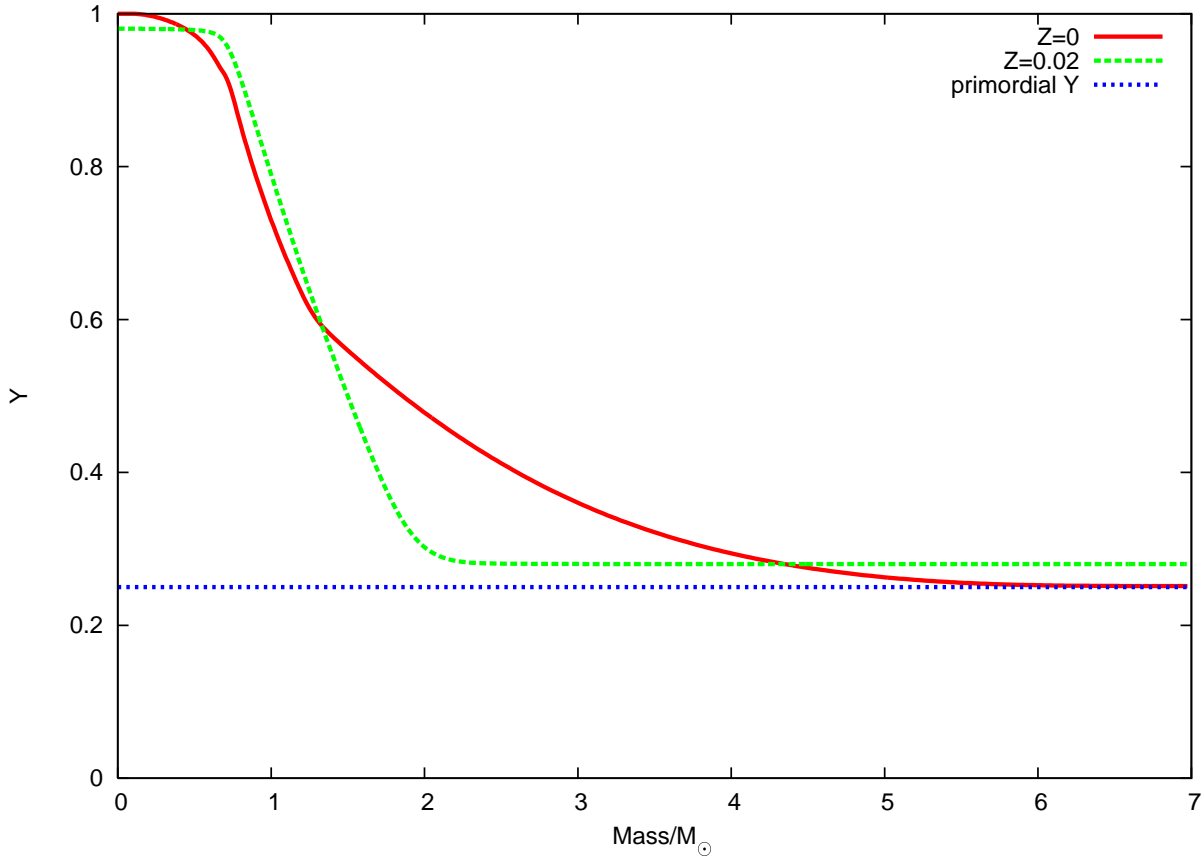}
\caption{Helium abundance by mass fraction profiles for solar and zero metallicity $7\,M_\odot$ models after hydrogen core burning. Notice that, for the $Z=0$ model, the helium abundance increases over a larger part of the star than for the solar metallicity model.}
\label {fig:7He}
\end{figure}

\shortcite{Chieffi} describe the central H and He burning of $4-7\,M_\odot$ stars. Our models agree with the characteristics they find. The convective core is much smaller than in metal-rich stars of similar mass. In our models it vanishes when the central mass fraction of H is 0.54, compared to 0.5 in their models. Our pp-chain-driven convective core is slightly smaller then theirs, while our CNO-cycle-driven convective core is slightly bigger than theirs. Secondly, as seen in Figure \ref{fig:7He}, the He abundance increases noticeably in $70-80\,$ percent of the star by mass, in agreement with \shortcite{Chieffi}. This is because the pp-chain is less temperature dependent than the CNO cycle. The helium abundance doubles out to about $2\,M_\odot$. The temperature gradient in the star is shallower than it would be in a higher metallicity star of the same mass. Consequently, hydrogen is burned in a more extended region, so the helium abundance increases over a larger part of the star.

\shortcite*{Siess} describe the AGB phase of a $7\,M_\odot$ zero-metallicity star too. We compare our core masses and surface abundances at this early AGB phase with theirs and the work of \shortcite{Chieffi} in Table~1. Our models are richer in nitrogen by a factor of ten and less abundant in carbon and oxygen by about a factor of ten and thirty respectively than those of Siess et al (2002). This suggests that their second dredge up is much deeper than ours and may even dip into the helium shell. This would also explain the drop in surface helium abundances for their  $7\,M_\odot$ model compared to their own $5\,M_\odot$ model.

\begin{figure}
\includegraphics[ width=\textwidth]{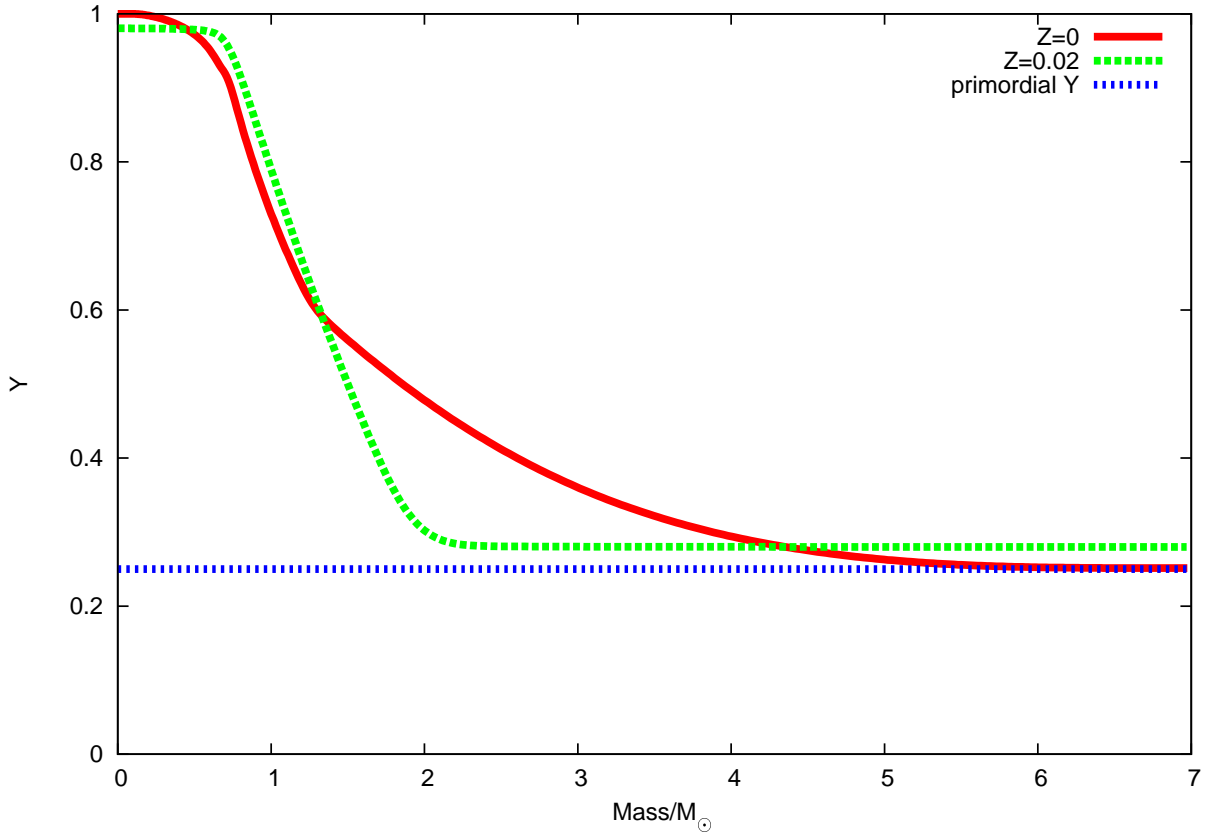}
\caption{Mass fraction profile of the $7\,M_\odot$ model at the end of central He burning. Note the significant amount of carbon and nitrogen produced at the hydrogen burning shell.}      
\label {fig:7mass}
\end{figure}

Our results are closer to those of  \shortcite{Chieffi}. The carbon and oxygen surface abundances agree very well for the $7\,M_\odot$ models. One significant difference is that our $7\,M_\odot$ model has a surface abundance of nitrogen about ten times higher. Comparison with their mass fraction profiles reveals that  nitrogen is produced in a more extended mass range in our model, (compare Fig.~\ref{fig:7mass} with Fig. 7 in their paper) so more of it is made and hence dredged up to the surface. This is because the hydrogen shell is thicker and the CNO cycle proceeds over a wider range of mass in our models.

\begin{table}
\begin{center}
\begin{tabular}[t] {c c c c c c}
\hline
Physical quantity & This work & \shortcite{Siess} & \shortcite{Chieffi}\\
\hline
core mass/$M_\odot$ & 1.0301 & 1.0268 & 0.9875\\
${}^{4}\rm{He}$ & 0.3821 &  0.3764 &  0.369 \\
${}^{12}\rm{C}$ & $2.64 \times 10^{-6}$ & $2.41 \times 10^{-5} $ &  $2.08 \times 10^{-6} $\\
${}^{14}\rm{N}$ & $2.28 \times 10^{-8}$ & $1.44 \times  10^{-9}$ & $1.59 \times  10^{-9} $\\
${}^{16}\rm{O}$ & $3.85 \times 10^{-9}$ & $9.39 \times 10^{-8}$ & $2.88 \times 10^{-9}$\\
\hline
\end{tabular}
\caption{Comparison of core masses and surface abundances by mass fractions of the  $7\,M_\odot$ zero-metallicity early AGB models.}
\end{center}
\end{table}

\section{The Late AGB phase of the $7\,M_\odot$ model}
We have continued the evolution of the $7\,M_\odot$ model through its entire thermally pulsing (TP) phase without any mass loss.  There are 590 thermal pulses in $1.1 \times 10^{5}\, \rm yr$. The interpulse period is about 700 yr for the first few pulses and decreases to about 100 yr at the end of the TP phase. The core mass is $1.04\,M_\odot$ when the first thermal pulse starts. Our interpulse periods are significantly shorter than the $2{,}700\,\rm{yr}$ found by \shortcite{Gilpons}. Also, our core mass is bigger than their model without overshooting by $0.08\,M_\odot$.  In fact their model with overshooting has an interpulse period and core mass much closer to ours. The maximum helium luminosity of our model never exceeds $10^{5.5}\,L_\odot$. The pulses are too weak to lead to any third dredge-up, in agreement with the work of \shortcite{Gilpons}. Carbon previously brought to the surface during second dredge-up has been converted to nitrogen by hot-bottom burning \citep{IbenHBB}. At this point, the surface carbon and nitrogen abundances are $5.5 \times 10^{-7}$ and  $2.7 \times 10^{-6}$ by mass. In the absence of third dredge-up, hot-bottom burning has reduced the surface carbon abundance during the TP-AGB nor have any $s$-process elements been brought up to the surface. 

The issue of third dredge-up in AGB stars has been a contentious one for some time. It has been postulated that third dredge-up only happens in stars above a certain critical metallicity \citep{Komiya}. While our models agree with the work of \shortcite{Gilpons}, who find that third dredge-up is absent, both \shortcite{Chieffi} and \shortcite{Siess} find that third dredge up does occur in their zero-metallicity intermediate mass AGB stars. \shortcite{Chieffi} treated the convective boundaries according to the prescription of \shortcite{Herwig}, who use a  mixing scheme so efficient that the composition discontinuity between the two burning shells is smoothed out. This seems to indicate the efficiency of third dredge-up depends on the treatment of convection and the inclusion of extra-mixing mechanisms such as convective overshooting. However, \shortcite{Gilpons} find that the total amount of mass dredged up is very small even when overshooting is included. Prescriptions that overshoot into the processed core generally raise the metallicity and make the behaviour more like that of stars of higher metallicity that do undergo deep third dredge up \citep*{Stancliffe,Stancliffe05}. For example, in recent models of super-AGB stars, \shortcite{Doherty} find that a $9.5\,M_\odot$ star has a dredge-up efficiency $\lambda$ of 0.7. Dredge up efficiency is defined by the amount of H-exhausted core matter mixed into the envelope divided by the amount of core growth during interpulse period. However, \shortcite{Siess2} find no third dredge up. As they highlighted, the occurrence (or not) of third dredge-up depends sensitively on how one treats the convective boundaries, as well as whether one includes additional mixing mechanisms. The \textsc{stars} code uses  an arithmetic scheme for determining the diffusion coefficient for the mixing. It has typically given deeper dredge-up than found in other codes \citep*{Stancliffe,Stancliffe05}, so it is significant that we find no dredge-up in these models, when others do. It may be that we do not find TDUP because we do not apply extensive extra-mixing and we shall investigate this in future work.

\begin{figure}
\includegraphics[angle=270, width=\columnwidth]{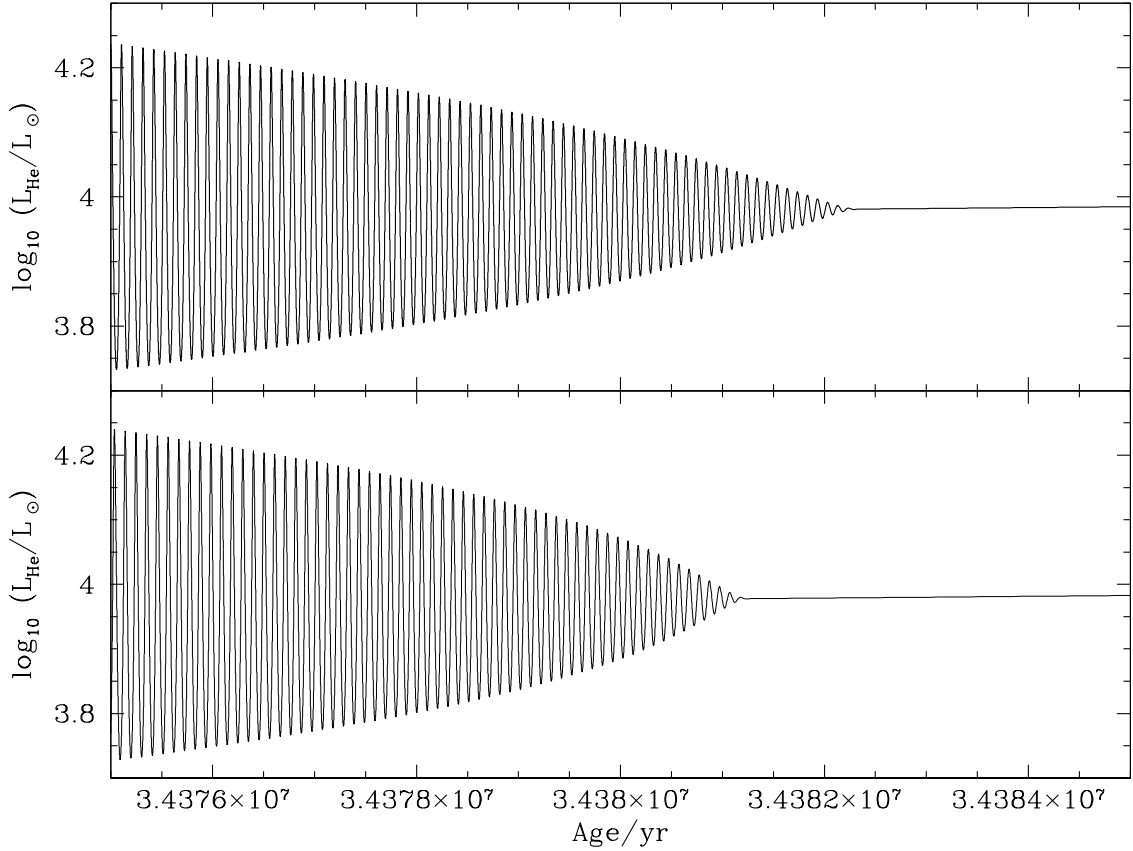}
\caption{The end of the thermal pulses of the star for the $7\,M_\odot$ model. The pulses grow weaker and eventually stop as indicated by the variation of helium luminosity. Top: Model without mass loss. Bottom: Model with Reimers' mass loss. The difference between the two models is very small.}     
\label{fig:pulses}
\end{figure}

Because helium burning already proceeds at a relatively high rate in the hotter burning shells during the interpulse period, the jump in helium luminosity during pulses is small compared to their higher metallicity counterparts. Later pulses are weaker and they cease altogether when the core mass reaches $1.1\,M_\odot$ (see Figure \ref{fig:pulses}). The star then enters a quiescent evolutionary phase for about $1.8 \times 10^{5}\,\rm{yr}$ while the hydrogen and helium burning shells grow outward without any thermal pulses. The star reaches carbon ignition after $2.9 \times 10^{5}\,\rm{yr}$, much faster than estimated by \shortcite{Gilpons} at $1.2 \times 10^{6}\,\rm{yr}$. This is because their estimate was based on the core growth rate of the first few pulses. However, although our core growth rate is $4.7 \times 10^{-7}\,M_\odot\,\rm{yr}^{-1}$, which agrees with their rate for the first few pulses, it increases with time and is $2.0 \times 10^{-6}\,M_\odot\,\rm{yr}^{-1}$ when it explodes. 

We have compared a $7\,M_\odot$ solar metallicity model with a  $5\,M_\odot$ zero-metallicity model with the same core mass in order to explain the weak thermal pulses of zero metallicity stars. We did not compare directly with the $7\,M_\odot$ zero-metallicity because its core mass is much larger than that of a $7\,M_\odot$ solar metallicity model. We find that the zero-metallicity star has a much thinner intershell, helium-rich layer both in terms of mass and radius. There is less helium to be burnt during the thermal pulses, so they are much weaker and the interpulse period is much shorter. Because of the shorter interpulse period, the helium shell does not cool down as much as the higher metallicity model, so the temperature in the helium shell increases after each pulse. This is in contrast to the solar metallicity stars where the helium shell temperature during the interpulse goes down after each pulse. Eventually, the temperature becomes hot enough that helium burning can proceed smoothly without any pulses. This is because helium burning is much less temperature sensitive at higher temperatures. At $T=10^{8}K$, the $3\alpha$ reaction rate is proportional to $T^{40}$ while at  $T=2\times 10^{8}K$ it is proportional to $T^{18.5}$ \citep* [cf. the stability criterion of][]{Yoon}. The hydrogen burning shell is much hotter and hence closer to the core in the zero-metallicity star. The higher temperature can cause earlier ignition of the pulse and hence the thinner intershell and helium burning shell. The hotter temperature can be attributed to the lower metallicity content of these stars, because of lower CNO abundances in the burning shell and lower opacity at the surface of the stars.

\begin{figure}
\includegraphics[ angle=270, width=\columnwidth]{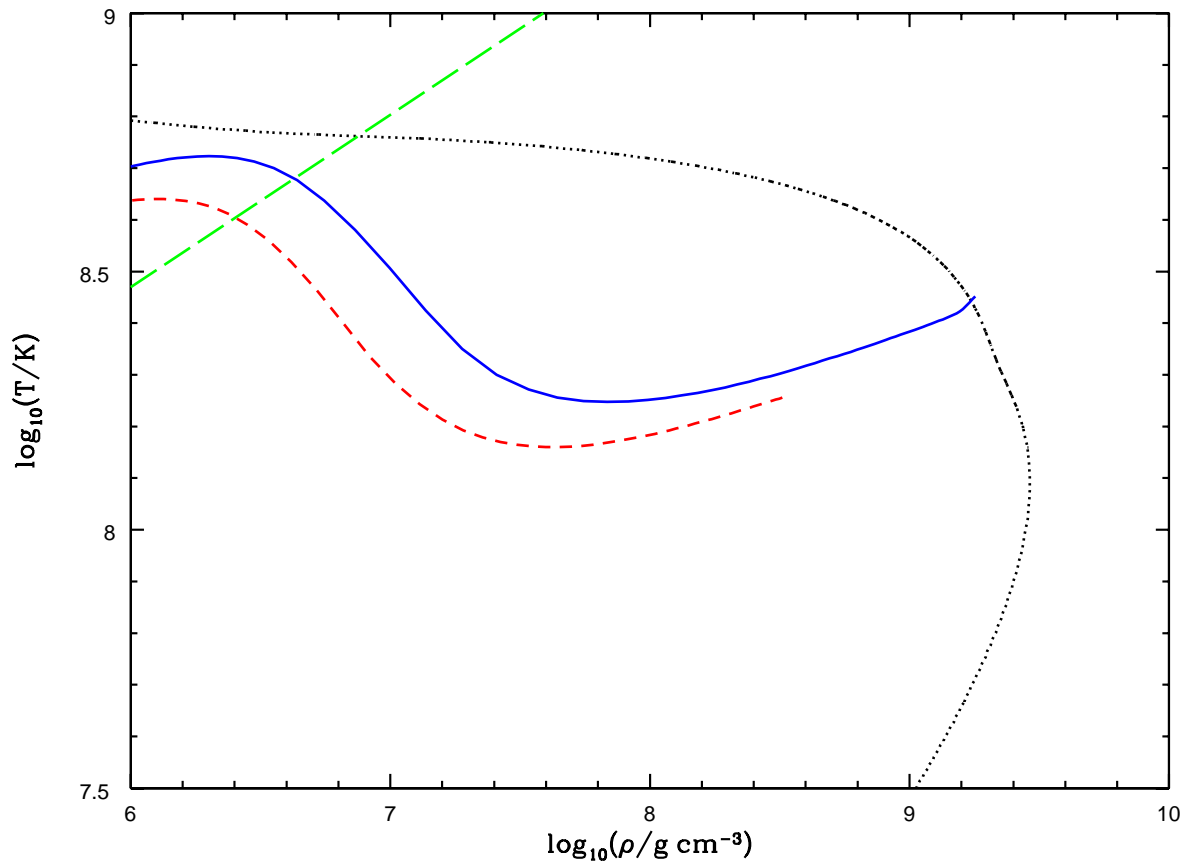}
\caption{Interior of the $7.0 \,M_\odot$ model. The dotted line indicates where carbon burning produces energy faster than it is lost via neutrinos. We add the internal structure profile with the core at the highest density and the burning shell at the highest temperature. The dashed line is at a slightly earlier time and the solid line is at the end of the evolution. Carbon ignites degenerately at the centre before the shell is hot enough to ignite carbon.  The long-dash line at the top left hand corner is the boundary between non-degenerate and degenerate regions.}
\label{fig:ignition}
\end{figure}

Carbon ignition occurs at the centre under degenerate conditions when the core mass reaches $1.36\,M_\odot$. The carbon luminosity rises rapidly and is soon followed by the breakdown of the evolution code because of the thermonuclear runaway.  We plot a carbon ignition curve as described by \shortcite*{Martin} in Fig.~\ref{fig:ignition}. Above and to the right of the solid line of the figure we have the right conditions for carbon ignition to drive a thermonuclear runaway. On the same axes, we plot the evolution of the internal temperature against density for the models leading up to ignition. Carbon ignites at the centre of the degenerate core because of the high density. Fig.~\ref{fig:ignition} shows that the core ignites carbon degenerately at the centre before any other part of the star. In particular, the burning shell is not hot enough. The subsequent thermonuclear runaway (similar to a type Ia supernova) releases sufficient energy to blow the whole star apart. Despite the envelope, we would expect the explosion mechanism of this star to be very similar to a Type Ia supernova, so we can estimate the nucleosynthetic yield from the exploding core. According to the deflagration model with zero metallicity (W70) of \shortcite{Iwamoto}, the nucleosynthesis products of a supernovae type Ia would be $5.08 \times 10^{-2}\,M_\odot$ of  ${}^{12}\rm{C}$, $3.31 \times 10^{-8}\,M_\odot$ of ${}^{14}\rm{N}$ and $0.133\,M_\odot$ of ${}^{16}\rm{O}$. As described by \shortcite*{Lau}, the yield of nitrogen from the envelope is of the order of $10^{-5}\,M_\odot$ and for carbon $10^{-6}\,M_\odot$ with an even lower oxygen yield, so we can conclude that the carbon and oxygen yields from the envelope are insignificant compared to the supernova yields. In the deflagration model nitrogen is mainly released by the envelope. However, if the explosion mechanism were delayed detonation the nitrogen yield of the core would be about $2\times10^{-4}\,M_\odot$ (WDD1, WDD2 or WDD3 of \citealt{Iwamoto}). In this case, the explosive yield would be higher than the yield from the convective envelope and may be the source of nitrogen for the N-enhanced stars described by \shortcite{Spite}. The carbon yield drops to about $10^{-2}\,M_\odot$ and the oxygen yield to about $7\times10^{-2}\,M_\odot$. In both scenarios, the two biggest yields are silicon and iron. The ${}^{28}\rm{Si}$ yield is $0.142\,M_\odot$ to $0.272\,M_\odot$ and ${}^{56}\rm{Fe}$ is $5.87\,M_\odot$ to $0.695\,M_\odot$. Such stars are important for the iron contribution in the early Universe. The composition of material ejected by the supernova can be very different to that released during binary interaction, as described by \shortcite{Lau} when the envelope is lost before the ignition of carbon. 

In the above estimates, we have ignored the nucleosynthesis that may take place in the envelope during the explosion. Unlike a Type Ia supernovae which does not have a hydrogen-rich envelope, this star could produce extra nucleosynthesis during its explosion just as a Type II supernova does when a shock wave sweeps through the envelope. The presence of an envelope also makes fallback a possibility. The ejecta can then be enriched by $\alpha$-capture isotopes and neutron-processed isotopes. If the star is in a binary system, r-processed elements formed during explosion can pollute the companion star. This may be a source of the double $r$/$s$-processed enriched halo stars, as suggested by \shortcite{Zijlstra}, even though $s$-process elements are not brought to the surface in our models .

\section{Mass-loss rate and the fate of the $7\,M_\odot$ model}

We have also evolved a $7\,M_\odot$ model with Reimers' mass loss. The evolution of this model is almost identical to that of the one without mass loss. Minor differences are that the thermal pulses stop about $1{,}000\,\rm{yr}$ earlier and the helium luminosity is generally lower, but by less than 0.1 dex, for the model with mass loss (see figure~\ref{fig:pulses}). The Reimers'  mass-loss rate  \citep{Reimersformula} is
 
\begin{equation}
\dot{M}_{\rm{R}}= - 4.0\times10^{-13} \eta \frac {(L/L_\odot)(R/R_\odot)} {(M/M_\odot)} M_\odot \,\rm{{yr}}^{-1},
\end{equation}
where $L$ is the luminosity of the star, $M$ its mass and $R$ its radius. We uses $\eta = 1$, which is likely to be an overestimate of the mass loss in low-metallicity stars. The maximum mass-loss rate, at the end of evolution, when the luminosity and radius are highest, is $8.1 \times 10^{-6}\,M_\odot\,\rm{{yr}}^{-1}$. The total mass lost from the star since the first thermal pulse is then about  $1.0 \,M_\odot$ and the average mass-loss rate is $3.4 \times 10^{-6}\,M_\odot\,\rm{{yr}}^{-1}$.  We have also estimated the mass-loss rate caused by the onset of radial pulsations \citep{VWML}. Because the star is much more compact than higher metallicity stars of similar mass, $R/R_\odot$ never exceeds $10^{2.7}$, so the maximum pulsation period is about 350 d. This period is significantly below the period for the onset of the superwind phase discussed by  \shortcite{VWML}. Therefore the mass-loss rate driven by pulsations is much lower than that of Reimers' prescription. 

The Reimers' rate was originally calibrated with M supergiants and so has no observational support for AGB stars so we briefly consider other mass-loss prescriptions such those of \shortcite{Schroder} and \shortcite{Blocker}, which give higher mass-loss rates. The formula given by \shortcite{Schroder} is 

\begin{equation}
\dot{M}_{\rm{SC}}= \left(\frac {T_{\rm{eff}}} {4,000\rm{K}}\right)^{3.5} \times \left(1+ \frac{g_\odot} {4,300g}\right) \dot{M}_{\rm{R}},
\end{equation}
where $T_{\rm{eff}}$ is the effective temperature of the star, $g$ is its surface gravity and ${g_\odot}$ is the surface gravity of the Sun. This prescription gives a mass-loss rate that is 10.8 times that of the Reimers' prescription but still applies to non-pulsating giants and so may not be applicable to AGB stars.

\shortcite{Blocker} gives the mass-loss rate as:

\begin{equation}
\dot{M}_{\rm{B}}= - 4.83\times10^{-9} \frac {{L}^{2.7}} {{M}^{2.1}} \dot{M}_{\rm{R}}.
\end{equation}
For solar metallicity stars, this gives consistently larger rates during the TP-AGB phases (e.g. \citealt{Gallart}) so, based on the numbers of luminous lithium rich AGB stars in the Magellanic Clouds, \shortcite{Ventura} suggested the use of $\eta=0.02$ for the Reimers' mass-loss rate when using this prescription. With this modification, the mass-loss rate is about 15 times the standard Reimers' rate and the timescale for the loss of the envelope is only $1.1 \times 10^{5}$ years. Both these formulae give a mass-loss timescale shorter than the evolution time and the envelope could be lost slightly before carbon ignition. However, all our above estimates have neglected the effect of metallicity. We can apply the commonly used scaling, suggested by \shortcite{Nugis} for hot stars, that

\begin{equation}
\dot{M}(Z) =\dot{M}(Z_\odot) \left(\frac {Z} {Z_\odot}\right)^{0.5} , 
\end{equation}
where $\dot{M}(Z_\odot)$ is the mass-loss rate for solar metallicity and $Z$ is the surface metallicity. The scaling arises from the assumption that stellar winds are line driven. With lower surface opacity at lower metallicity there are weaker winds. However, while there is general agreement that mass loss falls with metallicity, there is a range of suggested values for the exponent. For example, \shortcite{Vink} suggested $\dot{M}(Z) =\dot{M}(Z_\odot)(\frac {Z} {Z_\odot})^{0.64} $ for B supergiants.  Based on equation (4), as shown in table 2, all the mass-loss timescales are significantly longer than the actual evolution time. Our estimated mass-loss timescale for \shortcite{Schroder} is roughly 10 times shorter than the timescale given in \shortcite{Gilpons}. Our surface abundances are very close to their model, so the differences in timescale are because the star grows larger as it evolves. 

The above scaling only applies to a radiation-driven wind and the relation comes from the fact that the efficiency of dust formation is reduced at low metallicity. However, as shown previously, the pulsation driven mass-loss rate is very low, even if the scaling does not apply. Also, different mass-loss prescriptions may not be scalable by this relationship.  Nevertheless, we plot the evolution time and mass-loss timescale with and without scaling in Fig.\ref{fig:massloss}. The timescales for the star to lose its envelope are all significantly longer than the actual time the star takes to evolve up to the point of carbon ignition with the scaling. Even without scaling, the timescales are similar to, or slightly shorter than the evolution time, so even a weak scaling with a much higher surface metallicity means that the star does not lose its envelope before exploding.  The limiting mass-loss rate is $1.9 \times 10^{5}\,M_\odot\rm{{yr}}^{-1}$. The unscaled Reimers' prescription gives a mass-loss rate much lower than the limiting rate. Unless the current prescriptions greatly underestimate the mass-loss rate from AGB stars, the fate of a $7\,M_\odot$ zero-metallicity star is to explode as type 1.5 supernova. However, whether a star at low metallicity can reach such a mass loss rate is not clear yet, due to lack of observational data.

\section{Mass-loss rate and the fate of the $5\,M_\odot$ model}
\begin{figure}
\includegraphics[ angle=270, width=\columnwidth]{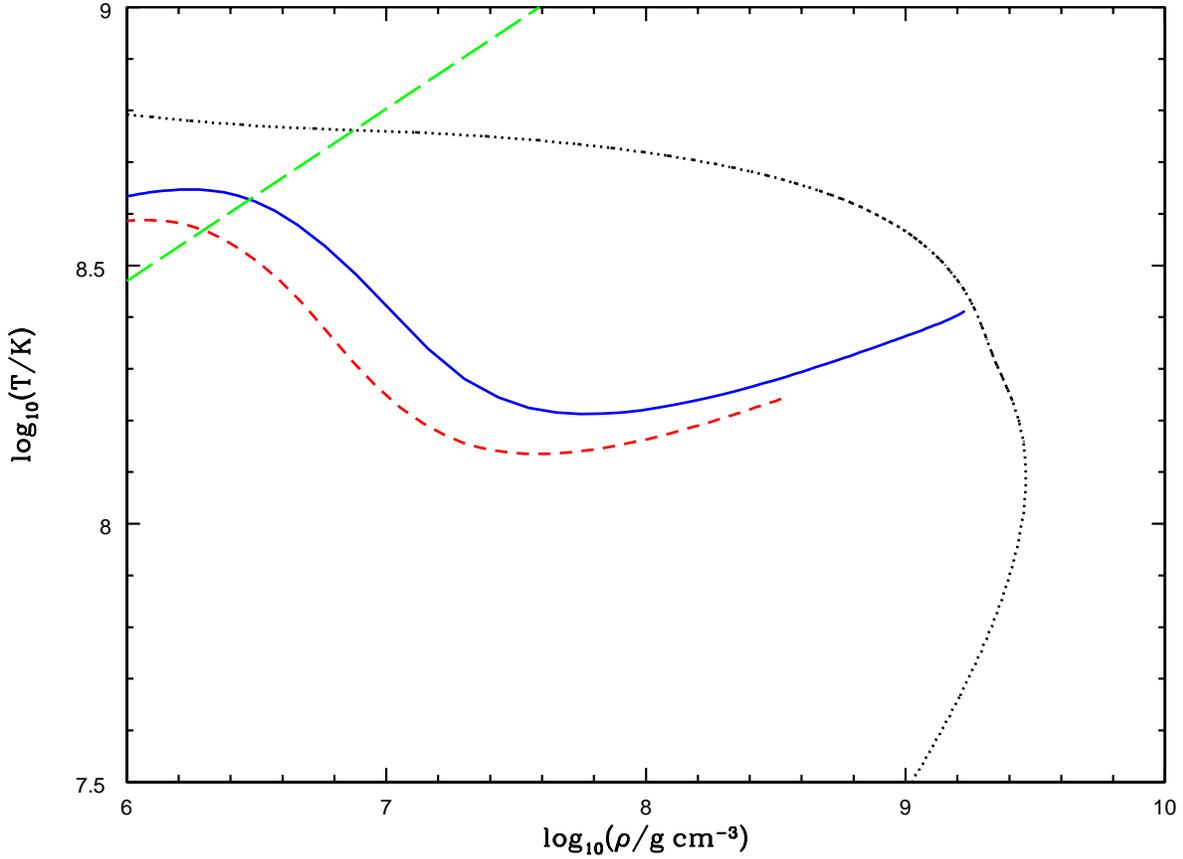}
\caption{Interior of the $5.0 \,M_\odot$ model. The dotted line indicates where carbon burning produces energy faster than it is lost via neutrinos. We add the internal structure profiles with the core at the highest density and the burning shell at highest temperature. The dashed line is at a slightly earlier time and the solid blue line is at the end of the evolution. Carbon ignites degenerately at the centre before the shell is hot enough to ignite carbon.  The long-dash line at the top left hand corner is the boundary between non-degenerate and degenerate regions.}
\label{fig:ignition2}
\end{figure}

\begin{figure}
\includegraphics[width =\columnwidth]{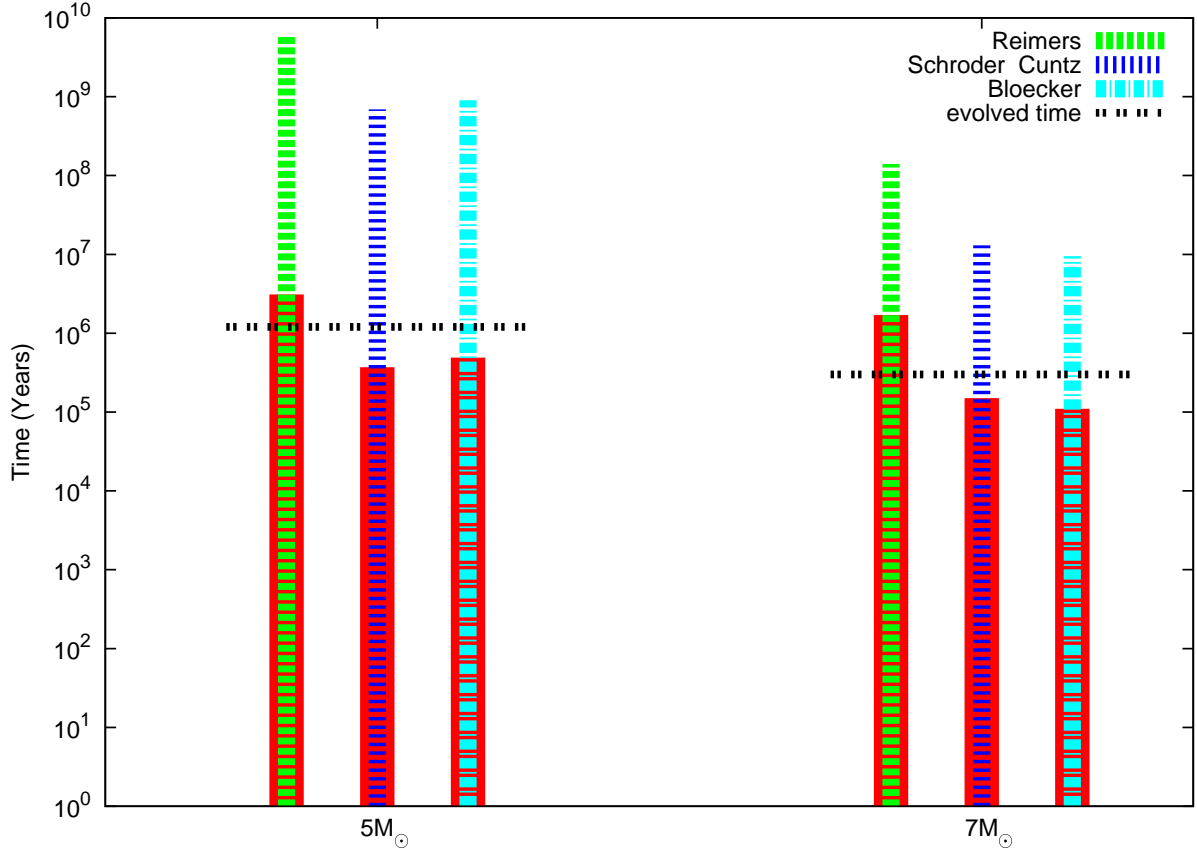}
\caption{Timescales to lose the envelope using different mass-loss rates compared with the actual evolution time. The solid bars are when no metallicity scaling is used while the broken bars give the longer timescales involved when the mass loss scaling $\dot{M}(Z) =\dot{M}(Z_\odot) (\frac {Z} {Z_\odot})^{0.5}$ is applied. It shows that only a small scaling is needed for the mass-loss timescales to be greater than the evolution time.}
\label{fig:massloss}
\end{figure}

We have also evolved a  $5\,M_\odot$ model with Reimers' mass loss with $\eta =1$. The evolution is very similar to that of the $7\,M_\odot$ star. When thermal pulses begin, the core mass is only $0.92\,M_\odot$ and pulses cease when the core has grown to $1.05\,M_\odot$. Like the $7\,M_\odot$ star, it also enters a quiescent phase until carbon ignites degenerately in the centre of the core when the core mass reaches $1.36\,M_\odot$ (see Fig. \ref{fig:ignition2}). The total time from the onset of thermal pulses to the onset of carbon ignition is $1.2 \times 10^{6}\,\rm{yr}$, much longer than the  $7\,M_\odot$ star because the whole thermally-pulsing AGB phases lasts much longer. The surface metallicity of the $5\,M_\odot$ model is much lower because the convective envelope does not reach as deep during second dredge-up. The surface CNO abundances by mass fraction near explosion are $2.3 \times 10^{-10}$, $5.1 \times 10^{-9}$ and $3.2 \times 10^{-11}$.

The total mass lost from the star is $1.4\,M_\odot$, with an average mass-loss rate of $1.1 \times 10^{6}\,M_\odot\rm{{yr}}^{-1}$. We have made a similar calculation of the mass-loss timescale based on different mass-loss rates (see Table 2). The limiting mass-loss rate is  $3.0 \times 10^{6}\,M_\odot\rm{{yr}}^{-1}$ so the possibility that the envelope is lost before carbon ignition is slightly higher because the core needs more time to grow. However, if we assume the mass-loss rate scales with metallicity, the mass-loss timescale is again much longer than the evolution time. As in the case of the $7\,M_\odot$ star,  we can be fairly confident that carbon ignition and the following supernova do occur, unless the effect of overshooting or other extra mixing mechanisms such as rotation increase the surface metallicity and hence the mass-loss rate.

\begin{table}
\begin{center}
\begin{tabular}[t] {c c c c c c c}
\hline
&&&&Time to lose envelope &\\
$M_{\rm{ZAMS}}$/$M_\odot$ & Metallicity Scaling & evolved time/yr  &  \shortcite{Reimers}/yr & \shortcite{Schroder}/yr &  \shortcite{Blocker}/yr               \\
\hline
5 & No &$1.2 \times 10^{6}$ & $3.1 \times 10^{6} $ &  $3.7 \times 10^{5} $& $4.9 \times 10^{5} $\\
7 & No &$3.0 \times 10^{5}$ & $1.7 \times  10^{6}$ & $1.5 \times  10^{5} $ & $1.1 \times  10^{5} $\\
\hline
5 & Yes &$1.2 \times 10^{6}$ & $6.0 \times 10^{9} $ &  $6.9 \times 10^{8} $& $9.0 \times 10^{8} $\\
7 & Yes &$3.0 \times 10^{5}$ & $1.4 \times  10^{8}$ & $1.3 \times  10^{7} $ & $9.5 \times  10^{6} $\\

\hline
\end{tabular}
\caption{Timescales associated with loss of the envelope for different mass-loss rates compared with the actual evolution time. The top section has no metallicity-scaling of the mass loss is used. The bottom section uses the relation $\dot{M}(Z) =\dot{M}(Z_\odot) (\frac {Z} {Z_\odot})^{0.5}$.  }
\end{center}
\end{table}

\section{conclusion}
We have shown that the fate of high-mass AGB primordial stars is to ignite carbon degenerately at their centres and explode as supernovae with behaviour similar to a Type Ia but with a hydrogen-rich envelope because the mass-loss rate is low for these stars. Such supernovae enrich the early Universe with metals such as iron, nickel and carbon. Whether this occurs depends on the mass-loss rate from the star. A high mass-loss rate can cause the star to lose its envelope before carbon ignition and end its evolution as a white dwarf. For the  $7\,M_\odot$ star, the critical average mass-loss rate is $1.9 \times 10^{5}\,M_\odot\rm{{yr}}^{-1}$ while for the $5\,M_\odot$, it is $3.0 \times 10^{6}\,M_\odot\rm{{yr}}^{-1}$. So far there is no observational support for any of the proposed mass-loss rates at low metallicity. In order to be certain that these stars explode, we require the mass-loss rate of these stars to be less than about one-third of the solar rate. Because of the low surface metallicity of these objects, the mass-loss rate should be low enough unless the surface metallicity of the stars is substantially increased by extra mixing or the current mass-loss metallicity scaling is very wrong.

We have also shown that for extremely metal poor stars, the strength of thermal pulses is weak and there is a lack of third dredge-up. Eventually the thermal pulses disappear and the core growth rate is much faster. This is important for a supernova type 1.5 to occur because the core can grow much faster than the time it takes for star to lose its envelope. The lack of third dredge-up also has important implications for the contribution of AGB stars to the chemistry of the early Universe, particularly for $s$-process isotopes which we are not produced without it. Further models of the TP-AGB phases of very low metallicity stars are necessary to determine the lowest metallicity at which third dredge up begins.
\section{Acknowledgments}
We thank Albert Zijlstra for his useful comments. HBL thanks PPARC for his Dorothy Hodgkin Scholarship. RJS and CAT thank Churchill College for their Fellowships.

\bibliography{HBL}

\label{lastpage}

\end{document}